# Progress in organic single-crystal field-effect transistors

Ignacio Gutiérrez Lezama and Alberto F. Morpurgo

DPMC and GAP, Université de Genéve, 24 quai Ernest Ansermet, CH-1211 Geneva, Switzerland

Research on organic thin-film transistors tends to focus on improvements in device performance, but very little is understood about the ultimate limits of these devices, the microscopic physical mechanisms responsible for their limitations, and, more generally, the intrinsic transport properties of organic semiconductors. These topics are now being investigated through the study of transport in organic transistors realized using molecular single crystals of unprecedented chemical purity and structural quality. These studies are elucidating detailed microscopic aspects of the physics of organic semiconductors and corresponding devices and have also led to unforeseen high values for carrier mobility in these materials. Here, we discuss developments in this area and present a brief outlook on future goals that have now come into experimental reach.

**Introduction organic, crystal, semiconducting electronic material, devices**

The development of organic field-effect transistors (FETs) based on conjugated molecules or polymers is driven by applications in the field of plastic electronics,[1–7] in which molecular semiconductors are used to produce large-area, low-cost, flexible electronic devices. These devices also provide an ideal setting for the development of a basic understanding of the microscopic physical mechanisms associated with charge transport in organic semiconductors. For instance, it is not well understood why some molecules lead to charge-carrier mobilities that are much larger than those for others. When compared to inorganic semiconductors such as silicon or III–V compounds, it seems clear that technology based on organic semiconductors would vastly benefit from a much more systematic, basic understanding of the electronic properties of these materials.

Investigating the intrinsic transport properties of organic semiconductors and their interfaces, which determine the transistor performance, requires materials of the highest quality to minimize extrinsic effects. Research groups have been pursuing the study of FETs based on single crystals of organic conjugated molecules,[8–18] which are now setting benchmarks for the performance of organic FETs. Single-crystal devices have led to observations of new physical phenomena and to the exploration of molecular materials that are pushing the limits of organic electronics beyond what had been initially foreseen. Here we provide a short introduction to the field, with special focus on the interplay between the transport properties of organic semiconductors and the physics of organic devices. In doing so, we also discuss the microscopic mechanisms that determine charge carrier motion in these systems.

**Fabrication of organic single-crystal transistors**

A variety of techniques are used to realize FETs[19] based on organic single crystals. Crystals can be grown directly on substrates, for instance by letting a drop-casted solution containing molecules evaporate,[20] or by seeding (vapor-phase) crystal growth at controlled locations.[21] The most common fabrication technique, however, separates crystal growth and transistor assembly.[19] A common strategy relies on manual lamination (see **Figure 1**) of organic crystals grown from vapor phase onto a substrate, in which the gate, source, and drain contacts are fabricated prior to lamination.[8,10,11] One can choose between a solid conducting substrate (acting as a gate) coated with a dielectric of choice,[8,10,11,18,19] or elastomer stamps[22,23] covered with a metal layer, molded to form the source and drain electrodes, and a recessed gate (so that air or vacuum acts as a dielectric). These latter devices show the largest mobility values and highest quality, as manifested by the observation of a band-like temperature dependence of the carrier mobility[24] (see below for more details on band-like transport). Techniques of this type have been applied to a broad variety of different molecular crystals to investigate both *p*- and *n*-channel devices[23–27] (see Figure 1).

Separating crystal growth and device assembly represents an advantage that can hardly be over-emphasized. Laminating previously grown crystals ensures that the quality of the organic material is always the same, which facilitates the correlation of device structure to performance. This is not possible for thin-film transistors, where the performance of FETs based on the same molecule frequently varies due to, for example, variations in thin-film morphology. In the investigation of the contact resistance, for instance, molecules deposited on the metal electrodes and on the dielectric (i.e., the channel) pack differently—resulting in grain boundaries,[28–30] whose effects on transport could not be separated experimentally from those of the metal/semiconductor interface. Organic single-crystal FETs realized by lamination offer high device-to-device reproducibility and have enabled the systematic investigation of mobility anisotropy,[23,31,32] the influence of the gate dielectric on the mobility,[33,34] bias-dependent contact resistance,[35,36] and charge-transfer at metal-organic interfaces.[37] Even though lamination on $SiO_2$ wafers or polydimethylsiloxane (PDMS) stamps is typically performed in ambient conditions, experiments indicate that interfacial contamination does not notably affect the electronic properties of the resultant single-crystal devices that show a surprising reproducibility.[38] This is likely due to the hydrophobic nature of smooth surfaces of molecular crystals with no grain boundaries or other defects that usually facilitate physisorption and chemisorption of contaminants.

**Device electrostatics**

In semiconducting devices, electrostatics determines the local density of charge carriers and plays a dominant role[39] that needs to be understood in order to correctly describe and interpret transport experiments. This point is illustrated by the behavior of space-charge limited current,[40] where seemingly minor perturbations (e.g., small densities of surface traps) can drastically affect the electrostatic profile.[41,42] From the viewpoint of electrostatics, even though it makes no difference if a transistor is realized using an organic or a conventional semiconductor (e.g., silicon), some important differences remain when it comes to the underlying device physics. In contrast to commonly used inorganic semiconductors, organic semiconductors are undoped, and one may wonder whether this difference invalidates conventional theory established for silicon FETs. The formation of a Schottky barrier at a metal/semiconductor interface, for instance, originates from the electrostatic profile of the bands in the semiconductor that is determined by the dopant density,[39] and it has long remained an open question whether the conventional Schottky theory[39,43,44] applies to organic materials. To settle such questions, it is important to discriminate—in experiments—between new physical phenomena that may occur in organic semiconductors and other unrelated effects. In organic thin-film transistors, the situation is more complicated because the presence of grain boundaries affects the electrostatics, often leading to either irreproducible or unconventional device behavior, which is at odds with established theory.[28–30] Organic single-crystal devices, on the contrary, are rather immune to these shortcomings.

A series of experiments on rubrene single-crystal FETs has examined the degree to which device electrostatics conform to the behavior expected for conventional inorganic transistors. These include the study of short channel FETs[35,36] and Schottky-gated transistors[45] (so-called MESFETs). Owing to the high carrier mobility, the resistance of short-channel devices (see **Figure 2**a) is dominated by the metal/semiconductor interfaces and can be modeled as two oppositely biased Schottky diodes.[35] The measurements directly give information about the Schottky barrier height and its electric field dependence (i.e., the Schottky effect). In devices with copper electrodes,[36] it was found that the conventional theory of transport through a Schottky barrier[39,43,44] reproduces the data quantitatively (Figure 2b–c), with physically sensible and internally consistent parameters. In the same short-channel transistors used to investigate the Schottky barrier, the analysis is further supported by the study of the length and bias dependence of the threshold voltage[37] associated with the charge transferred from the metal to the organic semiconductor. Quantitative agreement (Figure 2d) between experiments and theoretical estimates was found when using the system parameters extracted from the study of the contact resistance and of the MESFET devices (e.g., Schottky barrier height, density of unintentional dopants present in rubrene), without the need to introduce any additional free parameter.[37] It can therefore be concluded that for FETs realized on organic materials of sufficiently high structural quality, the device electrostatics are correctly described by the conventional theory established for inorganic transistors.

**Microscopic physics of organic semiconductors**

The fact that the electrostatics of organic and inorganic single-crystal FETs can be described by similar mathematical models does not mean that the underlying microscopic physics of charge transport is the same in the two cases. Most differences originate from the fact that in organic semiconductors, the constituent molecules are held together by weak van der Waals forces.[46,47] The electronic bandwidths associated with the highest occupied molecular orbital and the lowest unoccupied molecular orbital are much smaller than in common inorganic semiconductors (a few hundred meV in molecular crystals and approximately 10 eV in inorganic semiconductors). A small bandwidth in organic semiconductors implies that charge carriers are very sensitive to interactions[46,47] (e.g., with molecular vibrations, other carriers, and disorder). Indeed, the relative impact of such interactions is determined by the ratio of their strength to the relevant bandwidth.

*Band-like transport*

An important breakthrough enabled by organic single-crystal transistors is the observation[23–25] of the so-called band-like transport regime[48,49] at finite carrier density, with signatures of an increase in mobility with decreasing temperature,[24–26,50] observation of the Hall effect,[25,27,51] and an anisotropic mobility.[23,31,32] A Rutgers University (Podzorov and Gershenson)/University of Illinois (Rogers) collaboration[24] first observed these phenomena in rubrene single-crystal FETs. Band-like transport has been reported in a number of other compounds, for instance in tetramethyl tetraselena fulvalene (TMTSF),[26] 2,7-dioctyl[1]benzothieno[3,2-*b*][1] benzothiophene ($C_8$-BTBT),[50] and *N,N′*-bis(n-alkyl)-(1,7 and 1,6)-dicyanoperylene-3,4:9,10-bis(dicarboximide) (PDIF-$CN_2$)[27] single-crystal devices, both for holes and electrons (**Figure 3**). For holes,[34] mobility values as high as 20 cm$^2$/Vs at room temperature, increasing to approximately 40 cm$^2$/Vs at 150–200 K, have been observed; for electrons,[27] $\mu$ ~5 cm$^2$/Vs at room temperature has been observed, reaching ~10 cm$^2$/Vs at 200 K.

These achievements were not foreseen when organic single-crystal FET research started. Room temperature mobility values of 20 cm$^2$/Vs[24,34] largely exceed what was reported long ago ($\mu$ ~ 1 cm$^2$/Vs) by Norbert Karl in textbook time-of-flight (TOF) measurements[52] on zone-refined single crystals of different conjugated molecules. Nevertheless, in FETs the mobility is invariably found to decrease upon lowering temperature below 150–200 K, whereas in the best zone-refined organic crystals, the TOF mobility increases down to liquid helium temperature (reaching values of several 100s cm$^2$/Vs). This apparent inconsistency remains to be understood. Apart from possible differences in the two measurement techniques—TOF probes optically excited carriers whose energy is larger than that of carriers responsible for transport in FETs—the different low-temperature behavior is likely an extrinsic effect due to contamination of the crystal surface, affecting carriers in the transistor channel but not in TOF experiments (which probe the bulk).

Microscopically, the nature of band-like transport in organic FETs is not yet understood, although, theoretical progress has been made.[48,49,53–56] An important step was to realize that at room temperature, molecular motion (rotations and vibrations) leads to large fluctuations in the hopping integrals –the matrix elements of the charge carrier Hamiltonian between states at neighboring sites– which are of the order of the equilibrium values.[54] Charge carriers respond on a time scale much faster than the molecular motion and experience the random, thermally induced molecular configurations as very strong disorder causing Anderson localization (i.e., a complete localization of the carrier wave-function due to quantum interference ). On a longer time scale, the molecular configuration changes, and the localized carriers diffuse "following" the molecular motion. The mobility increases with lowering temperature because at lower $T$, the amplitude of the molecular motion decreases, and the localization length increases.[49,54–56] Since in the temperature range investigated the localization length is never much larger than the lattice spacing, such a regime is different from true band-transport, expected to occur only at much lower $T$ (in actual materials, however, extrinsic disorder takes over, causing "static" carrier localization and a steep decrease of the field-effect mobility). While there is consensus that this scenario is physically correct for high-purity organic single-crystal FETs, the problem remains difficult to treat theoretically because several important energy scales have a comparable magnitude:[47,49] the bandwidth, its fluctuations due to molecular motion, temperature, and disorder. It is important to develop theoretical schemes enabling controlled approximations to systematically address this transport regime.

*Charge carriers and their dielectric environment*

Considerable progress has been made in understanding the interaction between charge carriers in the FET channel and the nearby gate dielectric.[33,34,57–61] Apart from chemical groups present at the dielectric surface acting as traps (very important, especially for electron transport[62]), the electrical polarizability of the gate insulator plays a key role.[33,34,57] If the polarizability is large and originates from slow (compared to the characteristic electronic times) degrees of freedom, charge carriers couple strongly to the polarization cloud that they themselves induce in the dielectric (in simple terms, their image charge). This coupling amplifies the trend of the charge carriers toward localization, and it is observed experimentally[34] that a crossover occurs from band-like transport to thermally activated hopping upon increasing the polarizability of the dielectric (**Figure 4**). In the strong coupling regime, the observations can be quantitatively described in terms of polaron formation-a quasiparticle formed by the coupling of a charge carrier to the electrical polarizability of the surrounding medium,- using a well-defined microscopic theory that could be extended to explain the high carrier density regime, with non-negligible Coulomb interactions between carriers.[61,63] Coulomb interactions cause the FET conductivity to saturate at large carrier density, because mutual Coulomb repulsion suppresses polaron hopping. Organic single-crystal FETs have therefore enabled controlled investigations of polaron physics at low and high carrier densities (see also the article by Xie and Frisbie in this issue).

The coupling to the gate dielectric remains important even when the polarizability is small, as in polymeric insulators. True interfacial polarons do not form in this case. The interaction is better described in terms of the phenomenological concept of "dipolar disorder" originally introduced by Veres,[57,58] who studied hopping transport in devices based on disordered conjugated polymers. The monomers forming the polymer chains inside the gate insulator possess electric dipoles that are randomly oriented. They generate a spatially fluctuating electrostatic potential in the transistor channel, which broadens the energy distribution of the states in the organic semiconductor. As a result, the density of states at the Fermi energy decreases, leading to a decrease in the hopping probability and in the carrier mobility. This concept has recently been applied to single crystals, where transport occurs in the band-like regime:[27] dipolar disorder deepens the distribution of states in the band tail (essentially the shallow traps) in the organic semiconductor, resulting also in suppression of the field-effect mobility. An interesting recent development comes from the realization that in devices with a suspended channel, where the organic crystal is not in contact with a gate dielectric, it is the organic crystal itself that determines the "dielectric environment" experienced by the charge carriers accumulated in the FET channel (i.e., the bulk of the crystal plays the role of a gate dielectric). It is therefore important to understand how charge carriers couple to the electrical polarizability of the molecular planes adjacent to the FET channel. It has been suggested that this coupling can be minimized by specific molecular packing in the organic crystals and the structures of the constituent molecules,[27] resulting in favorable conditions for the occurrence of band-like transport.

*Where does the disorder come from?*

As compared to thin-film FETs, the level of disorder in the best suspended organic single-crystal transistors is negligible. To turn on a single-crystal device at room temperature, a density of carriers of only $\sim 10^{10}$ cm$^{-2}$ needs to be accumulated. Nevertheless, disorder still creates problems that become apparent upon lowering $T$, again because of the narrow bands of the organic semiconductors with their large associated density of states. At room temperature, the Fermi level is located in the disorder-induced tail of states at the band edge[26,64] but is sufficiently close to the band edge such that a large number of thermally excited carriers populate states with a large localization length (responsible for band-like transport). As temperature is lowered, the distance between the Fermi level and the band edge eventually exceeds $kT$ ($k$ is the Boltzmann constant), and charge carriers occupy only strongly localized states.[26] With its large density of states, a band tail 20–30 meV deep can host a large carrier density, and it is hard to shift the Fermi level close to the band edge by applying the gate voltage. Reducing the magnitude of disorder is therefore essential to investigate the intrinsic transport properties below 100 K. Another viable strategy relies on the so-called charge-transfer interfaces[38,65,66] between two different organic crystals—where charge is transferred from one material surface to the other—to bring the chemical potential inside the band (i.e., outside the tail).

A difficulty is our limited understanding of the dominant mechanisms causing disorder. Residual chemical impurities can generate states in the material bandgap, acting as deep traps. At the concentration levels estimated in the best materials (~$10^{14}$ cm$^{-3}$),[45] these states can be completely filled by applying a very small gate voltage and would not pose major problems. However, molecular impurities can deform the crystal lattice or, if charged, generate potential fluctuations, causing the formation of band tails behaving as shallow traps. The investigation of band-like transport in TMTSF single-crystal FETs suggests that a correlation between deep traps, whose concentration is estimated from the shift of the threshold voltage with temperature, and shallow traps, which affect the mobility, is present.[26] Structural disorder, such as dislocations or mechanical stress, can also play a role. Mechanical stress is likely induced during the transistor assembly process, with compression or stretching of the crystal causing local changes in the band width (because the hopping integral—hence the bandwidth—depends exponentially on the intermolecular distance), which leads to the formation of "pockets" responsible for charge trapping. Finally, the most pronounced band-like transport is observed in PDMS stamp devices[24,27,34] (Figure 3), where the crystal surface is exposed to ambient conditions and adsorbates that can introduce disorder. As a term of comparison, even on as-fabricated suspended graphene, potential fluctuations larger than 20–30 meV due to adsorbates are present[67] and can only be eliminated by annealing the devices in vacuum. It may not be a coincidence that in most organic single-crystal FETs, in which band-like transport is observed, the characteristic depth of the disorder-induced band tail is inferred from the transport data to be also 20–30 meV. Annealing organic single-crystal devices would be desirable, but it is unclear whether the delicate organic crystals (and elastomer stamps) can withstand the required elevated temperatures.

**Conclusions and outlook**

Progress in the area of organic single-crystal field-effect transistors has been considerable, setting new benchmarks for device performance, introducing new materials, and deepening our fundamental knowledge. Understanding the microscopic mechanisms explaining why some materials exhibit band-like transport and others do not, as well as the possibility to push band-like transport to lower temperatures, are targets for current research. These targets are now within reach because of the increasingly larger number of different organic semiconductors available that exhibit band-like transport and reasonable (~1 $cm^2$/Vs) mobility values at sub-100 K temperatures, enabling systematic comparative studies. Much will depend on our ability to understand the origin of—and ability to minimize—disorder, for which the broader class of molecules now available is also advantageous. As has been the case for a long time in the field of organic electronics, progress is steady, which gives good reason to be optimistic.

**Acknowledgments**

We are grateful to H. Alves, R. de Boer, I.N. Hulea, N. Iosad, A.S. Molinari, C.L. Mulder, M. Nakano, N. Minder, S. Russo, A.I. Stassen, and H. Xie for their contributions to different aspects of our research on organic single-crystal FETs through the years. We would also like to thank our collaborators S.

Ciuchi, A. Facchetti, S. Fratini, M.E. Gerhenson, Y. Iwasa, V. Podzorov, T. Takenobu, and J. Takeya for their contributions to our work. We acknowledge financial support from the SNF, NCCR MaNEP, and NEDO during the period in which the manuscript was prepared.

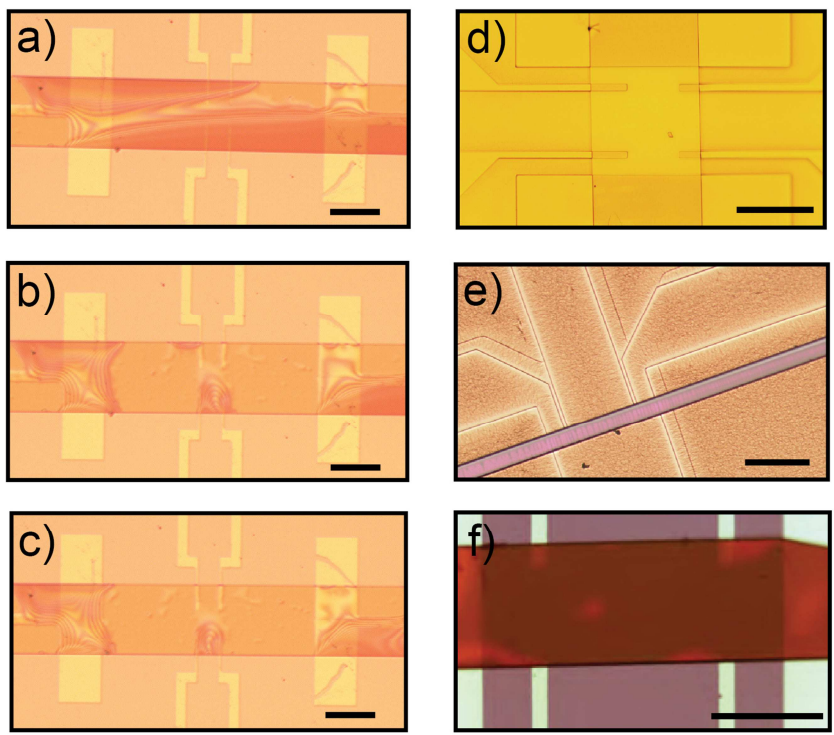

Figure 1. (a–c) Optical micrographs illustrating the time evolution of the spontaneous adhesion of a rubrene single crystal to an SiO$_2$ substrate during lamination.[19] (d–f) Optical micrographs offering a top view of several single-crystal field-effect transistors. (d) Rubrene singe crystal laminated on top of a polydimethylsiloxane stamp covered with a gold layer. The gate electrode is recessed, hence vacuum acts as the dielectric.[34] (e) Similar device as in d) made with a tetramethyl tetraselena fulvalene single crystal.[26] (f) N,N′-bis(n-alkyl)-(1,7 and 1,6)-dicyanoperylene-3,4:9,10-bis(dicarboximide), n-type single-crystal laminated onto a cytop (amorphous fluoropolymer) film.[27] The scale bars in (a–e) are 200 μm; in(f) it is 100 μm.

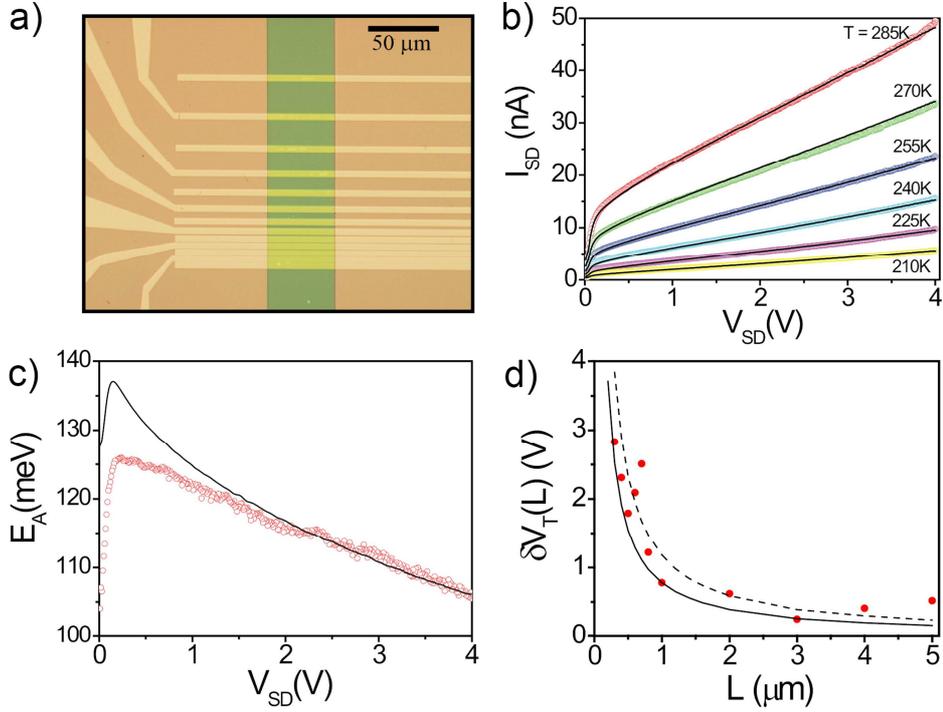

Figure 2. Transport through short-channel rubrene field-effect transistors with Cu electrodes. Notice in (a) how many different transistors can easily be realized on a same single crystal. (b) The source/drain current-voltage characteristics ($I_{SD}$–$V_{SD}$) (symbols) of these devices, measured at different temperatures ($T$), can be modeled in terms of two oppositely biased Schottky diodes, using the conventional Schottky theory (solid lines).[36] (c) The bias-dependence of the Schottky barrier height ($E_A$, circles) can also be reproduced using the same theory[36] (solid line). (d) Short-channel devices show a length-dependent threshold voltage shift ($\delta V_T (L)$, circles) due to charge-transfer from the metal contacts, which becomes non-negligible for submicron channel lengths ($L$).[37] The length dependence of the threshold voltage can be accounted for quantitatively using the known range of Schottky barrier heights, 0.13 eV (dashed line) and 0.15 eV (solid line) obtained from the study of the contact resistance performed on the same devices.

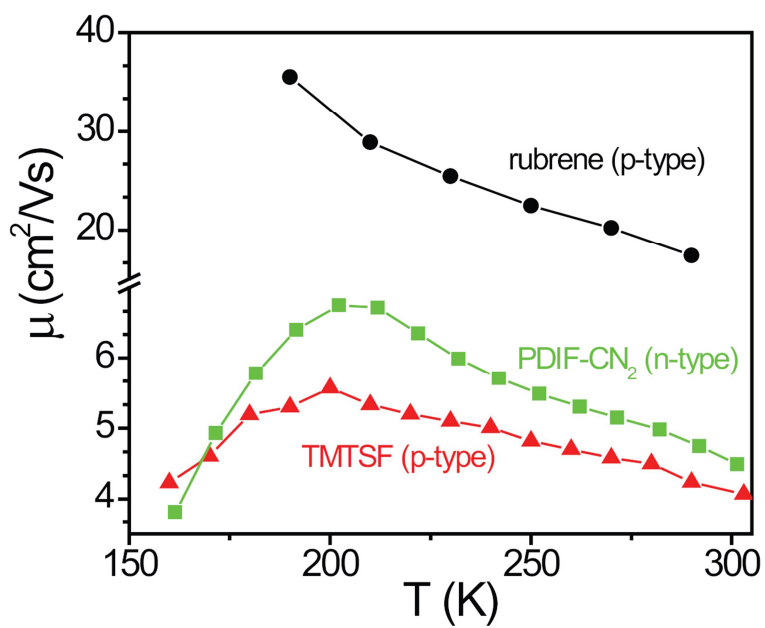

Figure 3. Band-like transport in single-crystal field-effect transistors laminated on top of gold-covered polydimethylsiloxane stamps (similar to those shown in Figure 1), in which single crystals of tetramethyl tetraselena fulvalene (TMTSF),[26] *N,N′*-bis(n-alkyl)-(1,7 and 1,6)-dicyanoperylene-3,4:9,10-bis(dicarboximide) (PDIF-CN$_2$),[27] and rubrene[34] are suspended above a gate electrode. At low temperatures, the mobility (μ) decreases as temperature (*T*) is lowered due to disorder-induced trapping of charge carriers (not shown for rubrene).

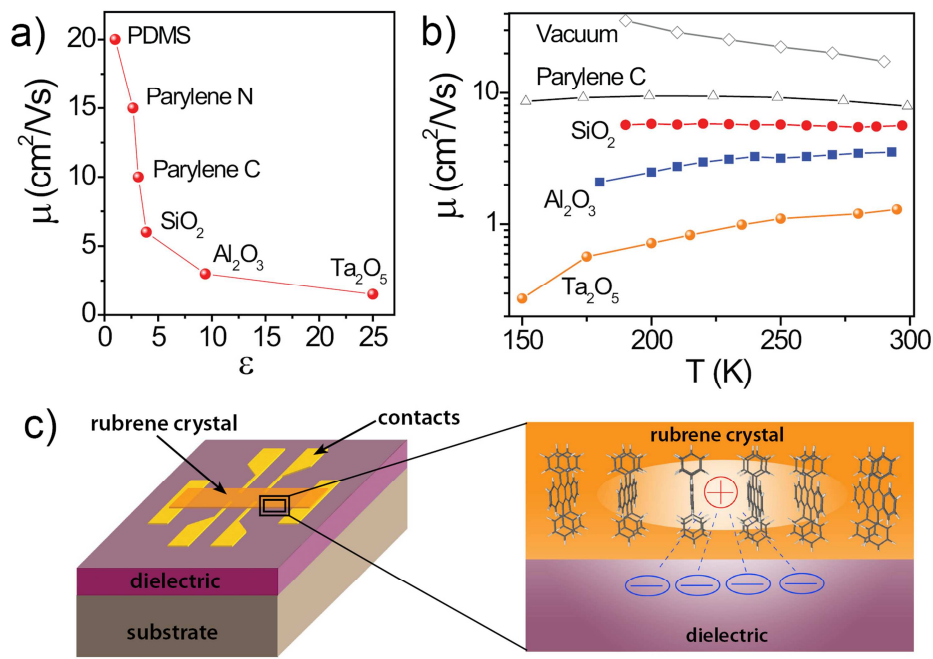

Figure 4. Influence of the gate dielectric on charge-carrier mobility (μ) and its temperature dependence (*T*), as measured on rubrene single crystals with different gate insulators.[33,34] (a) At room temperature, the carrier mobility decreases with increasing dielectric constant (ε), while a crossover from band-like transport to thermally activated hopping is observed in the temperature dependence, as shown in (b) by the decrease of the mobility with decreasing temperature in transistors with higher gate dielectric constants. Note: PDMS, polymethylsiloxane. (c) Schematic view of how holes accumulated in the transistor channel—at the organic/dielectric interface—polarize the gate dielectric[34].